\documentclass[aps,prl,reprint,showpacs,superscriptaddress,amsmath, nolongbibliography]{revtex4-1}

\usepackage{graphicx}
\usepackage{txfonts}
\usepackage{hyperref}
\usepackage{amsmath,bm}
\usepackage[utf8]{inputenc}

\newcommand{\vv} {{\bm v}}
\newcommand{\be}{\begin{equation}} 
\newcommand{\ee}{\end{equation}}  
\newcommand{\bea}{\begin{eqnarray}}
\newcommand{\eea}{\end{eqnarray}}

\newcommand{\cO}{{\cal O}}

\renewcommand{\v}[1]{\mathbf{#1}} 

\newcommand{\f}[2]{\dfrac{#1}{#2}}
\newcommand{\avg}[1]{\left\langle #1 \right\rangle} 

\newcommand{\re} {\mathrm e}

\graphicspath{{./images/}} 

\begin{document}
\title{Rare events statistics of random walks on networks: localization\\and other dynamical phase transitions}

\author{Caterina De Bacco}
\affiliation{LPTMS, Centre National de la Recherche Scientifique et Université Paris-Sud 11, 91405 Orsay Cedex, France.}
\author{Alberto Guggiola}
\affiliation{LPTENS, Unité Mixte de Recherche (UMR 8549) du CNRS et de l’ENS, associée à l’UPMC Université Paris 06, 24 Rue Lhomond, 75231 Paris Cedex 05, France.}
\author{Reimer K\"{u}hn}
\affiliation{Department of Mathematics, King's College London}
\author{Pierre Paga}
\affiliation{Department of Mathematics, King's College London}

\begin{abstract}
Rare event statistics for random walks on complex networks are investigated using the large deviations formalism. 
Within this formalism, rare events are realized as typical events in a suitably deformed path-ensemble, and their 
statistics can be studied in terms of spectral properties of a deformed Markov transition matrix. We observe two
different types of phase transition in such systems: (i) rare events which are singled out for sufficiently large 
values of the deformation parameter may correspond to {\em localized\/} modes of the deformed transition matrix;
(ii) ``mode-switching transitions" may occur as the deformation parameter is varied. Details depend on the nature
of the observable for which the rare event statistics is studied, as well as on the underlying graph ensemble. In 
the present letter we report on the statistics of the average degree of the nodes visited along a random walk 
trajectory in Erd\H{o}s-Rényi networks. Large deviations rate functions and localization properties are studied 
numerically. For observables of the type considered here, we also derive an analytical approximation for the 
Legendre transform of the large-deviations rate function, which is valid in the large connectivity limit. It is 
found to agree well with simulations.
\end{abstract}

\maketitle

Random walks are dynamical processes widely used to analyze, organize or perform important tasks on networks
such as searches \cite{adamic2001,guimera2002}, routing or data transport \cite{servetto2002, tadic2004,
tadic2004information}. Their popularity is due to their cheap implementation, as they rely only on local 
information, such as the state of the neighborhood of a given node of the network. This  ensures network 
scalability and allows fast data transmission without the need for large storage facilities at nodes, such as 
big routing tables in communication networks. These features make random walks an efficient tool to explore 
networks characterized by a high cost of information. Examples are sensor networks \cite{tian2005} where many signaling packets are needed to acquire wider networks status information. In peer-to-peer networks the absence of a central server storing file locations requires users to perform repeated local searches in order to find a file to download, and various random walk strategies have been proposed as a scalable method \cite{chawathe2003, lv2002,bisnik2005} in this context. Less attention has been paid to characterize rare events associated with random walks on networks. Yet the occurrence of a rare event can have severe consequences. In hide-and-seek games for instance \cite{sneppen2005}, rare events represent situations where the seeker finds either most (or unusually many) of the hidden targets, or conversely none (or unusually few). In the context of cyber-security, where one is concerned with worms and viruses performing random walks through a network, a rare event would correspond to a situation where unusually many sensible nodes are successfully attacked and infected, which may have catastrophic consequences for the integrity of an entire IT infrastructure. Characterizing the statistics of rare events for random walks in complex networks and its dependence on network topology is thus a problem of considerable technological importance. A variant of this problem was recently analyzed for biased random walks in complex networks \cite{kishore2012}. That paper addressed rare fluctuations in single node occupancy for an ensemble of independent (biased) walkers in the  stationary state of the system. By contrast, our interest here is in rare event statistics of {\em path averages\/}, or equivalently of time integrated variables. Rare event statistics of this type has been looked at for instance in the context of kinetically constrained models of glassy relaxation \cite{Garrah+09}; relations to constrained ensembles of trajectories were explored in \cite{JackSoll10} for Glauber dynamics in the 1d Ising chain. While these studies were primarily concerned with the use of large deviations theory as a tool to explore dynamical phase transitions in homogeneous systems, our focus here is on the interplay between rare event statistics and the heterogeneity of the underlying system. 

In the present Letter we use large deviations theory to study rare events statistics for path averages of 
observables associated with sites visited along trajectories of random walks.  Within this formalism, 
rare events are realized as typical events in a suitably deformed path-ensemble \cite{touchette2009, Garrah+09}. 
Their statistics can be studied in terms of spectral properties of a deformed version of the Markov transition 
matrix for the original random walk model, the relevant information being extracted from the algebraically 
largest eigenvalue of the deformed transition matrix. Such deformation 	 may direct random walks to subsets of
a network with vertices of either atypically high or atypically low coordination. It also amplifies the 
heterogeneity of transition matrix elements for large values of the deformation parameter and we observe 
that, as a consequence, the eigenvector corresponding to the largest eigenvalue of the deformed transition 
matrix may exhibit a {\em localization transition\/}, indicating that rare large fluctuations of path averages 
are typically realized by trajectories that remain localized on small subsets of the network. Within localized
phases, we also encounter a second type of dynamical phase transition related to {\em switching between modes\/} 
as the deformation parameter used to select rare events is varied. Our methods allow us to study the role that 
network topology and heterogeneity play in selecting these special paths, as well as to infer properties 
of paths actually selected to realize extreme events.

\paragraph{The model.} 
We consider a complex network with adjacency matrix $A$, with entries $a_{ij}=1$ if the edge $(ij)$ exists, 
$a_{ij}=0$ otherwise. The transition matrix $W$ of an unbiased random walk has entries $W_{ij}=a_{ij}/k_{j}$ 
where $k_{j}$ is the degree of node $j$ and $W_{ij}$ is the probability of a transition from $j$ to $i$. 

Writing $\v{i}_\ell = (i_0, i_1, \cdots,i_\ell)$ a path of length $\ell$, quantities of interest are empirical 
path-averages of the form 
\be
\hat \phi_\ell = \frac 1 \ell \sum_{t=1}^{\ell} \xi_{i_t}\ ,
\ee
where the $\xi_i$ are quenched random variables associated with the vertices $i=1,\dots,N$ of the graph, which 
could be independent of, be correlated with, or be deterministic functions of the degrees $k_i$ of the 
vertices. It is expected that the $\hat \phi_\ell $ are for large $\ell$ sharply peaked about their mean
\begin{equation}\label{eqn:phi}
\bar{\phi}_{\ell}=\frac{1}{\ell}\sum_{\v{i}_\ell} \, P(\v{i}_\ell) \, \sum_{t=1}^{\ell} \, \xi_{i_{t}} =
 \left\langle \frac{1}{\ell}\sum_{t=1}^{\ell} \xi_{i_{t}}\right\rangle
\end{equation}
where $P(\v{i}_l)$ denotes the probability of the path $\v{i}_l$.

The average (\ref{eqn:phi}) can be obtained from the {\em cumulant generating function\/} $\psi_{\ell}(s) = 
\ell^{-1}\ln \sum_{\v{i}_\ell} \, P(\v{i}_\ell) \, \re^{s\sum_{t=1}^{\ell} \xi_{i_{t}}}$ as $\bar{\phi}_{\ell}= 
\psi_\ell'(s)|_{s=0}$. Here, we are interested in rare events, for which the empirical averages $\hat \phi_\ell$ 
take values $\phi$ which differ significantly from their mean $\bar \phi_\ell$. Large deviations theory predicts 
that for $\ell \gg 1$ the probability density $P(\phi)$ for such an event scales exponentially with path-length 
$\ell$, $P(\phi) \sim \re^{-\ell I(\phi)}$, with a {\em rate function\/} $I(\phi)$ which, according to the 
G\"artner-Ellis theorem \cite{touchette2009} is obtained as a Legendre transform $I(\phi)=\sup_{s}\{ s\phi - 
\psi(s)\}$ of the limiting cumulant generating function $\psi(s) = \lim_{\ell\rightarrow \infty} \psi_{\ell}(s)$, 
provided that this limit exists and that it is differentiable. We shall see that the second condition may be
violated, and that the derivative $\psi'(s)$ may develop discontinuities at certain $s$-values, entailing that
we observe regions where $I(\phi)$ is strictly linear and only represents the convex hull of the true rate 
function \cite{touchette2009}.

In order to evaluate $\psi_{\ell}(s)$, we express path probabilities using the Markov transition matrix
$W$ and a distribution $\v{p_0}=(p_0(i_0))$ of initial conditions as $P(\v{i}_\ell)= \big[\prod_{t=1}^{\ell}
W_{i_{t}i_{t-1}}\big]\,p(i_{0})$, entailing that $\psi_\ell(s)$ can be evaluated in terms of a deformed 
transition matrix $W(s) =\big(\re^{s \xi_{i}}\,W_{ij}\big)$ as $\psi_{\ell}(s) =\ell^{-1} \ln 
\sum_{i_{\ell},i_{0}} [W^{\ell}(s)]_{i_{\ell}i_{0}}\, p(i_{0})$. Using a spectral decomposition of the 
deformed transition matrix one can write this as
\begin{equation}
\psi_{\ell}(s) = \ln \lambda_1 + \frac{1}{\ell} \ln \Bigg[(\v 1,\v v_1\big) \big(\v w_1,\v p_0\big)
+\sum_{\alpha (\ne 1)} \Bigg(\frac{\lambda_\alpha}{\lambda_1}\Bigg)^\ell (\v 1,\v v_\alpha\big) 
\big(\v w_\alpha,\v p_0\big) \Bigg]\ .
\end{equation}
Here the $\lambda_\alpha$ are eigenvalues of $W(s)$, the $\v v_{\alpha}$ and $\v w_\alpha$ are the 
corresponding right and left eigenvectors, $\v 1=(1,\dots,1)$, and the bracket notation $(\cdot,\cdot)$ 
is used to denote an inner product. Eigenvalues are taken to be sorted in decreasing order $\lambda_1 \ge 
|\lambda_2|\ge |\lambda_3| \dots \ge \lambda_N$, with the first inequality being a consequence of  the 
Perron-Frobenius theorem \cite{Gantmacher59}. This concludes the general framework. For the remainder
of this Letter, we will restrict our attention to the case where $\xi_i= f(k_i)$.

For long paths, the value of the cumulant generating function is dominated by the leading eigenvalue 
$\lambda_{1}=\lambda_{1}(s)$ of the transition matrix $W(s)$, so $\psi(s)=\log \lambda_{1}(s)$. 
In the $s=0$ case, the eigenvalue problem is trivial, as the column-stochasticity of the transition matrix yields 
a left eigenvector $w_i \equiv 1$ corresponding to the maximal eigenvalue $\lambda_1 = 1$. The associated right 
eigenvector is $v_i \propto k_i$. For nonzero $s$, such closed form expressions are in general not known.
Performing a direct matrix diagonalization is quite daunting for large system  sizes $N$, even if one exploits 
methods that calculates only the first eigenvalue \cite{lanczos}. Hence we are interested in fast viable 
approximations. Here we describe one such approximation expected to be valid for networks in which vertex
degrees are typically large. 

\paragraph{Degree-based approximation.}
We start by considering the left eigenvectors $\v{w}$ instead of the right eigenvectors, for which the 
eigenvalue equation can be written as
 \begin{equation}
	\lambda\, w_{j}=\frac{1}{k_{j}} \sum_{i \in \partial j} w_{i}\, \re^{s f(k_{i})}\ .
\label{linsys}
\end{equation}

This system of equations can be simplified by considering a degree-based approximation for the first eigenvector, 
where one assumes that the values of $w_{i}$ only depend on the degree of the node $i$: $w_{i}=w(k_{i})$. If the 
average degree is large enough and the degree distribution is not too heterogeneous, we can write the eigenvalue 
equation \eqref{linsys} by appeal to the law of large numbers as
 \begin{equation}
\lambda_{1}(s)\, w(k)= \sum_{k'}P(k'|k)\,w(k')\, e^{sf(k')}
\label{linsys-k}
 \end{equation}
where $P(k'|k)$ is the probability for the neighbor of a node of degree $k$ to have degree $k'$.

\begin{figure}[!ht]
\begin{center}
\includegraphics[width=0.45\textwidth]{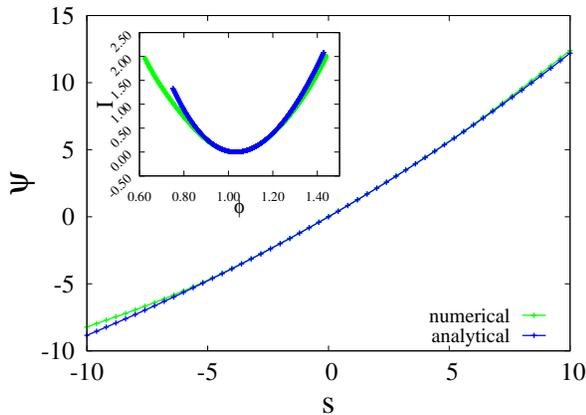}
\caption{(Colour online) Cumulant generating function $\psi(s)$ for ER networks with $c=30$ and 
$f(k_i)=k_{i}/c$, comparing the large-degree approximation (\ref{degaprox}) (blue line) with results of 
a numerical simulation (green line). The inset shows the corresponding rate functions.}\label{figdegaprox}
\end{center}
\end{figure}

In an Erd\H{o}s-Rényi (ER) ensemble \cite{ER}, and more generally in any configuration model ensemble, we 
have $P(k'|k) = P(k')\frac{k'}{\avg{k}}$. In this case the right-hand side of (\ref{linsys-k}) does not depend 
on $k$ and the $w(k)$ are in fact $k$-independent. The eigenvalue equation then simplifies to
\begin{equation}\label{degaprox}
\lambda_{1}(s) = \avg{\f{k}{\avg{k}}e^{sf(k)}}\ ,
\end{equation}
where the average is over the degree distribution $P(k)$. This approximation yields excellent results for 
large mean connectivities $c=\avg{k}$ on ER graphs, and more generally for configuration models without low 
degree nodes. This is illustrated in figure \ref{figdegaprox}, where we plot a comparison with numerical 
simulations for ER graphs with $c=30$. In figure \ref{figdegaprox} and throughout the remainder of the paper
simulation results are obtained as averages over 1000 samples.

\paragraph{Eigenvector localization.}
Because of the heterogeneity of the underlying system, one finds the random walk transition matrix to
exhibit localized states, both for fast and slow relaxation modes \cite{Ku15}, even in the undeformed
system, although the eigenvector corresponding to the largest eigenvalue (the equilibrium distribution)
is typically delocalized. However, given the nature of the deformed transition matrix, one expects the 
deformed random walk for large $|s|$ to be localized around vertices where $s f(k_i)$ is very large; hence
we anticipate that in the deformed system, even the eigenvector corresponding to the largest eigenvalue
{\em may\/} become localized for sufficiently large $|s|$. In order to investigate this effect quantitatively 
we look at the inverse participation ratio of the eigenvector corresponding to the largest eigenvalue 
$\lambda_{1}$ of $W(s)$. Denoting by $v_i$ its $i$-th component, we have
\be
{\rm IPR}[\v v] = \frac{\sum_i  v_i^4}{\Big[\sum_i v_i^2\Big]^2}
\ee
One expects ${\rm IPR}[\v v ] \sim N^{-1}$ for a delocalized vector, whereas  ${\rm IPR}[\v v ]= \cO(1)$
if $\v v$ is localized.

\paragraph{Results on random graphs.}
We performed numerical simulations to evaluate $\lambda_{1}(s)$ and the $ \rm IPR[\v v_1 (s)]$ for several types of 
network, defined by their random graph topology. In the present letter we restrict ourselves to discussing
results for ER  networks. We found that other network ensembles such as scale-free random graphs give 
qualitatively similar results; we will report on these in an extended version of this letter.

We looked at various examples for the function $f(k_i)$ but in the present letter we only report results
for the normalized degree $f(k_i) = k_{i}/c$; other deterministic types of degree-dependent functions 
exhibit similar behavior, thus focusing on the normalized degree is sufficient to capture the important aspects 
of this problem. We restrict our simulations to the largest (giant) component of the graphs, in order to 
prevent spurious effects of isolated nodes or small disconnected clusters (e.g. dimers) dominating $\lambda_{1}(s)$ 
and the IPR for negative $s$, as these would represent trivial instances of rare events, where a walker 
starts, and is thus stuck on a small disconnected component of the graph. From here on, the network size given must be understood as the size of the networks from which the giant component is extracted.

\begin{figure}[!ht]
\begin{center}
\includegraphics[width=0.45\textwidth]{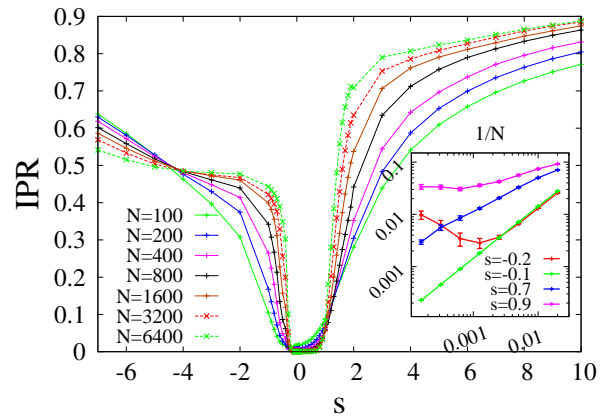}
\end{center}
\caption{$\rm IPR[\vv]$ as a function of the deformation parameter $s$ for ER graphs with $c=6$, and 
$f(k_i)= k_i/c$. The inset exhibits the $N^{-1}$-scaling of IPRs for 4 different values of the 
deformation parameter $s$, chosen in pairs on either side of {\em two\/} localization transitions, 
one at negative, and one at positive $s$.
}	
\label{IPR}
\end{figure}

Fig. \ref{IPR} shows the existence of two localized regimes for sufficiently large values of $|s|$, 
with IPRs on the localized side of both transitions increasing with system size. Results can be understood, 
as for large $|s|$ the deformed random walk is naturally attracted to the nodes with the largest (resp. 
smallest) degrees for positive (resp. negative) $s$. Thus for large negative $s$ the deformed walk tends 
to be concentrated at the end of the longest dangling chain, whereas for large positive $s$ it will be 
concentrated at the site with the largest available coordination. On an ER network where the large-degree 
tail of the degree distribution decays very fast, such a high degree vertex is likely to be connected to 
vertices whose degrees are lower, even significantly lower, than that of the highest degree vertex in the 
network, which leads to IPRs approaching 1 in the large $N$ limit. Conversely, for negative $s$, the 
deformed random walk will be attracted to the ends of dangling chains in the network, with the probability 
of escape from a chain decreasing with its length (with the length of the longest dangling chain increasing 
with system size). This can explain that IPRs initially saturate at 1/2 for large systems. Only 
upon further decreasing $s$ to more negative values will the asymmetry of the deformed transition matrices, 
to and away from the end of a dangling chain, induce that further weight of the dominant eigenvector to become 
concentrated on the end-site, leading to a further increase of the IPR.

\begin{figure}[!ht]
\begin{center}
\includegraphics[width=0.45\textwidth]{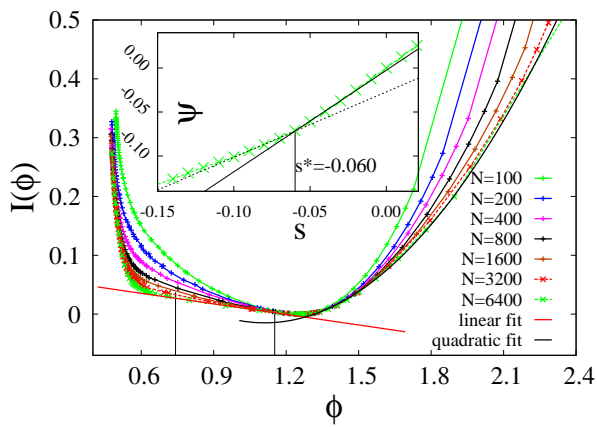}
\end{center}
\caption[]{Rate function $I(\phi)$ for ER graphs with $c=3$, and $f(k_i)=k_i/c$ for system sizes ranging 
from $N=100$ to $N=6400$. In the inset, we show $\psi (s)$ in the vicinity of the non-differentiable point. 
For the largest system size, a linear fit of the convex envelope of the left branch and a quadratic fit of 
the right branch of $I(\phi)$ are shown as well.}
\label{ratefunction}
\end{figure}

From the values of $\lambda_{1}(s)$ we also derived the large deviation rate functions for path averages
of the normalized degree $f(k_i)=k_{i}/c$, for various systems sizes and average connectivities.  In fig. 
\ref{ratefunction} we report $I(\phi)$ for an ER network at a low connectivity of $c=3$. While the right branch
of $I(\phi)$ is for large $N$ well approximated by a parabola, our results show the emergence of a linear region 
on the left branch, which becomes more pronounced as the system size is increased. This is a signature of a 
non-differentiable point of $\psi(s)$ at a point $s^{*}$ estimated to be at $s^{*}=-0.060 \pm 0.002$: at this 
point the G\"artner-Ellis theorem cannot be used to evaluate the rate function, and the linear branch only 
represents the convex envelope of the true $I(\phi)$ \cite{touchette2009}. The latter can either coincide with 
its convex envelope, or it can indeed be non-convex. However this information cannot be accessed by the theorem. 
The emergence of a jump-discontinuity in $\psi'(s)$ is due to a level crossing of the two largest eigenvalues, 
where the system switches between two modes that correspond to the largest eigenvalue on either side of $s^{*}$.
In finite systems the crossing is an `avoided crossing' due to level repulsion, but the two largest eigenvalues
become asymptotically degenerate at $s^{*}$ in the $N\to \infty$ limit, leading to a divergence of the 
correlation length $\xi(s) =[\ln(\lambda_1(s)/\lambda_2(s)]^{-1}$ at $s^{*}$, in close analogy with phenomenology 
of second order phase transitions, the divergence being logarithmic in $N$ in the present case.

\paragraph{Conclusions and future perspectives.}
In this Letter we have analyzed rare events statistics for path averages of observables associated with sites 
visited along random walk trajectories on complex networks. Results are obtained by looking at spectral properties
of suitably deformed transition matrices. The main outcome of our analysis is the possible emergence of two 
types of dynamical phase transitions in low mean degree systems: localization transitions which entail that large 
deviations from typical values of path averages may be realized by localized modes of a deformed transition matrix, and 
\textit{mode-switching transitions} signifying that the modes (eigenvectors) in terms of which large deviations are typically 
realized may switch as the deformation parameter $s$ and thus the actual scale of large deviations are varied. 
Results of numerical simulations consistently support these claims. We also developed an analytical approximation
valid for networks in which degrees are typically large.

Our work opens up the perspective to study a broad range of further interesting problems. On a technical level, 
one would want to implement more powerful techniques, such as derived in \cite{Kaba+10}, to obtain the largest 
eigenvalue in the present problem class for larger system sizes. Then there is clearly the need to systematically 
study the dependence of the phenomena reported here on the degree statistics, and on the nature of the observables 
for which path averages are looked at. We have gone some way in this direction and will report results in an 
extended version of the present paper. In particular one might wish to look at observables which, rather then being 
deterministic functions of the degree, are only statistically correlated with the degree, or at observables taking 
values on {\em edges between\/} nodes \cite{touchette2009, JackSoll10}. This could be of interest in applications 
such as traffic or information flows on networks subject to capacity constraints on edges. Moreover, given the nature 
of the mode-switching transition observed in the present letter, it is clearly conceivable that {\em several such 
transitions\/} could be observed in a single system, depending of course on the nature of the observables studied 
and on the topological properties of the underlying networks. Finally, critical phenomena associated with the 
localization transition and with mode-switching transitions also deserve further study. We believe that this list
could go on.

This work was supported by the Marie Curie Training Network NETADIS (FP7, grant 290038). 
	
\bibliographystyle{unsrt}
\bibliography{ver4}

\end{document}